\documentclass[a4paper]{article}
\usepackage{amsmath}
\usepackage{amssymb} 
\usepackage{amsthm}   
\usepackage{epsfig}
\usepackage{natbib}
\usepackage{a4wide}
\usepackage{url}
\usepackage[german,greek]{babel}
\usepackage{setspace}
\usepackage{cjhebrew}
\usepackage[hidelinks]{hyperref}
\usepackage{cjhebrew}

\begin{document}

\selectlanguage{german}
     \title{Der Stern von Bethlehem – Einige Theorien und Hintergr\"unde:
Von der altorientalischen Astrologie\\ \"uber Kepler zu Newton
\\ \large 
(nach einem Vortrag an der Theologischen Fakult\"at Fulda  am 12.12.2023)} 
        \author{Stephan F. Huckemann\thanks{ \url{https://www.stochastik.math.uni-goettingen.de/index.php?id=14&username=huckeman}}
        }
        \date{}
    
    \maketitle
    


   \section{Einf\"uhrung}
   
   Schon viele haben es unternommen, eine Abhandlung \"uber den \emph{Stern von Bethlehem} abzufassen, insbesondere \"uber Keplers Erkl\"arung als eine Sterngeburt, ausgel\"ost durch das Hinzutreten von Mars zu einer \emph{Gro{\ss}en Konjunktion} (so wird das Aufeinandertreffen von Jupiter und Saturn genannt), so wie er es im Oktober 1604 beobachtete. An vorderer Stelle ist hier sicher das Buch von \cite{ferrari1969stern}\footnote{mit neueren Auflagen bis 2003.} zu nennen. Es ist das Ziel meiner kurzen Abhandlung, auf einem hoffentlich gut verdaulichen Niveau in zugrundeliegende physikalische, astrologische und astronomische Vorstellungen der Antike um die Zeit von Christi Geburt einzuf\"uhren und ihre wesentlichen Weiterentwicklungen bis zur fr\"uhen Neuzeit, an denen Kepler ma{\ss}geblichen Anteil hatte, aufzuzeigen. Die Beitr\"age Keplers erm\"oglichten es ihm, erstaunlich exakt in das Jahr 7 vor Christus zur\"uckzurechnen, um dort eine dreifache Gro{\ss}e Konjunktion in den Fischen zu verorten, die dann von Mars komplettiert wurde. Dieses wird sowohl von der babylonischen \"Uberlieferung als auch von der modernen Astronomie best\"atigt. Ich hoffe, die nun folgende Besch\"aftigung mit den Hintergr\"unden ist f\"ur ein eher theologisch orientiertes Publikum hilfreich, auch weil die Kontroverse um \glqq geozentrische\grqq~und \glqq heliozentrische\grqq~Weltbilder (d.h. um Kopernikus und Galilei) gestreift (und in den Fu{\ss}noten weiter ausgef\"uhrt) wird, und weil sich aus dem Gesamtkontext Fragen an die Theologie ergeben, mit denen der Epilog schlie{\ss}t.
   
   \subsection{Gute Wissenschaft: Hypothesen und Grundlagen}
   
   Es ist das Kennzeichen guter Wissenschaft, bei aller Begeisterung f\"ur ein Forschungsprojekt, alle Hypothesen und Grundlagen, die zu einer Behauptung oder Entdeckung f\"uhren, sorgf\"altig anzugeben. 
   Als Beispiel, dass gute Wissenschaft kein Privileg der Neuzeit ist, mag folgender Satz gelten: 
   \begin{quote}{\it 
   Bei fast allen Fischen sind die Geschlechter unterschieden, aber es gibt einige Zweifel bei den Arten Eritrinus und Channa, denn alle stellten sich als schwanger heraus.}
   \end{quote}
   Popul\"arwissenschaftler w\"aren versucht, verk\"urzt zu behaupten, dass f\"ur diese beiden Fischarten nur Weibchen existierten. Dieser Satz stammt von Aristoteles ($\dagger$ 322 v. Chr.)\footnote{\citet[Buch VI, Kap. 12]{peck1943aristotle}.} und beweist, dass gute wissenschaftliche Arbeit, Zur\"uckhaltung und Offenheit der Antike gewiss nicht unbekannt waren.
   
   F\"ur den Stern von Bethlehem gibt es im Wesentlichen drei Hypothesen: Er war
   \begin{enumerate}
    \item eine kerygmatische Erz\"ahlung\footnote{so schreibt der katholische Alttestamentler \cite{steins2023wohin}: \emph{Der Stern als solcher sagt nichts und er beweist nichts, weil wir aus der von biblischen Motiven getr\"ankten Erz\"ahlung des Matth\"aus keinen Bericht machen k\"onnten, der dazu noch astronomisch ausgewertet werden k\"onnte. Das ist sachfremd, denn die Bibel spricht eine andere Sprache.} Der katholische Neutestamentler \cite{soding2023hintergrundinformationen} schreibt: \emph{F\"ur Matth\"aus ist wohl der Messiasstern leitend, den der heidnische Prophet Bileam vorhergesagt hat (Num 23,1; 24,17).}}, \emph{damit ihr glaubt} (Joh 20,30),
    \item eine atmosph\"arische Erscheinung, die lokal beobachtet wurde, oder
    \item er beschrieb ein astronomisch beobachtbares Ph\"anomen.
   \end{enumerate}
   Dieser Abhandlung liegt die dritte Hypothese zugrunde.
   
   \subsection{Biblischer Befund}
   
   Im zweiten Kapitel des Matth\"ausevangeliums lesen wir von \emph{Magiern aus den \"ostlichen (Gebieten)} ({\selectlanguage{greek}m'agoi >ap`o >anatol\~wn}), die nach Jerusalem an den Hof des K\"onigs Herodes kamen und den \emph{neugeborenen K\"onig der Juden} suchten: \glqq \emph{Wir hatten [...] den Stern im Osten (oder Aufgang?) gesehen} ({\selectlanguage{greek}e>'idomen [...] t`on >ast'era >en t\~h| >anatol\~h|})\grqq . Da erschrak Herodes und fragte, \emph{wann der Stern erschienen war}? Nun lernen die Magier, dass Bethlehem als Geburtsst\"atte in Frage kommt, das lag gut 10 km s\"ud-s\"udwestlich vom Palast des Herodes, der im heutigen Jerusalemer Armenierviertel verortet wird\footnote{z.B. \cite{peleg2019herod}.}. Laut biblischer Erz\"ahlung folgen die Magier dann dem Stern. Wenn er anfangs im S\"uden stand und sie ca. 2-3 Stunden auf ihn zugingen, dann war er ca. $30^\circ - 45^\circ$ nach Westen gewandert und f\"uhrte die Magier in einem Bogen also ziemlich genau nach Bethlehem.

   \subsection{Kandidaten f\"ur den Stern}
   
   Der astronomischen Hypothese folgend ergeben sich drei nat\"urliche Kandidaten f\"ur den Stern. Eine Supernova\footnote{einer \emph{stella nova} liegt \"ublicherweise eine Explosion eines Fixsterns zugrunde, \cite{bethe1990supernova} diskutiert diesbez\"ugliche physikalische Theorien von der Antike bis heute.} (derartige werden in koreanischen und chinesischen Quellen um die Jahre 5 und 4 v. Chr. berichtet\footnote{z.B. \cite{ze1966ancient}.}),  ein Komet (der Halleysche Komet war ca. 12 v. Chr. sichtbar\footnote{z.B. \cite{yoke1962ancient}, Cassius Dio Cocceianus ($\dagger$ ca. 235) berichtet dar\"uber in seinen \emph{Historiae Romanae} 54.29.8.}) und eine \emph{Konjunktion} (ein nahes Zusammentreffen zweier Himmelsk\"orper). Da Kometen und Supernovae Himmelsereignisse sind, die \"ublicherweise von einer breiten \"Offentlichkeit wahrgenommen und diskutiert werden, erschiene, wenn eines von diesen f\"ur den Stern st\"unde, die Frage des Herodes nach dem (ersten?) Beobachtungsdatum unverst\"andlich. Viel sinnvoller erscheint sie dagegen bez\"uglich einer sich ggf. wiederholenden Konjunktion, die subtilere Aufmerksamkeit erforderte. 
   
   Kometenbeobachtungen wurden auch im R\"omischen Reich niedergeschrieben, insbesondere die Beobachtung eines Kometen zu Caesars ($\dagger$ 44 v. Chr.)  Begr\"abnis\footnote{so berichtet Plinius d. \"Altere ($\dagger$ ca.  79) in seiner \emph{Naturalis Historia}, 2.34.}. Als Zeichen, dass Caesar zu den G\"ottern erhoben wurde, ergab sich so f\"ur Augustus ($\dagger$ 14) eine Begr\"undung f\"ur den anschlie{\ss}enden Kult des g\"ottlichen Kaisers\footnote{so ebenfalls Plinius ebd., vgl. auch die Diskussion in \cite{gurval1997caesar}. 
   }. Was liegt n\"aher, so eine exegetische Argumentationslinie\footnote{nach \cite{dieterich1902weisen} wurde das Sternmotiv in das Evangelium aufgenommen, nachdem die Beobachtung des Halleyschen Kometen im Jahr 66 mit einem zeitgleichen Besuch einer Delegation aus dem Morgenland in Rom, die Kaiser Nero anbetete, einen gro{\ss}en Eindruck auf die Zeitgenossen machte.}, als aus gleichem Grund zu Jesu Geburt einen Kometen zu verk\"unden? Obschon Origenes ($\dagger$ ca. 253)\footnote{Contra Celsum, 1,58.} durchaus die Kometenhypothese diskutiert, so entfaltete diese erst durch  Giotto di Bondone  kunstgeschichtliche Wirksamkeit, der offenbar angeregt durch die erneute Beobachtung des Halleyschen Kometen im Jahr 1301 diesen wohl erstmalig in ein Krippenfresko aufnahm\footnote{\cite{olson1987new} diskutieren dazu, wie sich die Kometenhypothese durch Origines und durch Apokryphe entwickelte.}. 
   
   \section{Keplers Stella Nova}
   Zu Weihnachten, im ersten Coronajahr (2020) konnte man abends kurz nach Sonnenuntergang im S\"udwesten 
   zwei helle Sterne dicht nebeneinander sehen: Saturn und Jupiter. Deren Zusammentreffen wird \emph{Gro{\ss}e Konjunktion} genannt. Eine \"ahnliche Konjunktion unmittelbar vor Sonnenaufgang beobachtete Kepler kurz vor Weihnachten 1603. Die Planeten bewegten sich dann langsam wieder auseinander und hatten bereits den gesamten Nachthimmel durchquert, als sich im Herbst 1604 Mars ihnen n\"aherte. Kurz darauf stand dort pl\"otzlich ein neuer, alles \"uberstrahlender Stern, eine auch noch etwa ein Jahr sp\"ater sichtbare Supernova. Nun nahm er zwar nicht an, dass die Konjunktion aller drei Planeten die \emph{stella nova} hervorgebracht h\"atte\footnote{Tycho Brahe hatte gemessen, dass unter der Hypothese, dass sich die Erde um die Sonne dreht, die Sterne nahezu unendlich weit entfernt seien (keine positive Parallaxe) und deshalb, da sie als kleine Lichtscheiben erschienen, absurd gro{\ss} seien m\"ussten. Kepler nahm die nahezu unendliche Entfernung hin, mit der Konsequenz, dass Planeten keinen Einfluss auf Sterne haben konnten \citep{graney2018mites}.}, aber er nahm an, dass Gott der Konjunktion als Erinnerungzeichen den neuen Stern beigef\"ugt habe\footnote{\citet[Cap. 27]{kepler1606destella}, eine umfassende Einordnung, inbesondere Keplers Zufallsbegriffs \emph{fortitudo (in sobrio sensu)} leistet \cite{recio2026incirco}.}. 
   
      
      \subsection{Astrologischer Hintergrund}
      Der Zeitpunkt der Sterngeburt war ein besonderer, denn laut zeitgen\"ossischen Vorstellungen handelte es sich gleichzeitig um die siebente Wiederkehr des \"Ubergangs vom Wasserdreieck zum Feuerdreieck und derartige \"Uberg\"ange h\"atten in der Vergangenheit stets mit bedeutsamen Weltereignissen korrespondiert\footnote{\citet{aston1970fiery}.}. Der erste \"Ubergang nach Erschaffung der Welt wurde in die Zeit Enochs verlegt, der zweite in Noahs Zeit, der dritte stand f\"ur Moses, der vierte verwies auf den Untergang des Reiches Israel, zum f\"unften wurde Jesus geboren, zum sechsten wurde Karl der Gro{\ss}e Kaiser und zum siebten sollte die Weltzeit nach insgesamt 6000 Jahren enden\footnote{ebd. S. 160.}. Die Periode von Gro{\ss}en Konjunktionen war wohlbekannt, etwa alle 20 Jahre ereignete sich eine und schritt im Tierkreis um etwa jeweils 4 Sternbilder zur\"uck. Somit durchliefen die Gro{\ss}en Konjunktionen etwa alle 60 Jahre 3 feste, jeweils um $120^\circ$ versetzte Tierkreiszeichen. Jedem dieser vier m\"oglichen Dreiecke war eines der vier Elemente (Erde, Wasser, Luft und Feuer) zugeordnet. Da die Periode tats\"achlich etwas kleiner als 4 Tierkreiszeichen war, kam es etwa alle 220 Jahre zu einem Wechsel und ein solcher fand 1603 statt. Die Welt ging aber nicht unter und so konnte Kepler auf der Grundlage seiner eigens entwickelten Himmelsdynamik sehr erfolgreich in das Jahr 7 v. Chr. zur\"uckrechnen, dort eine dreifache Gro{\ss}e Konjunktion in den Fischen mit anschlie{\ss}ender Marsbeteiligung verorten und somit im Analogieschluss postulieren, Gott habe auch damals einen neuen Stern hervorgebracht: den Stern von Bethlehem\footnote{\cite{kepler1614DeVeroAnno}.}.

      \subsection{Antike Physik und ptolem\"aische Himmelsmechanik}\label{scn:antik}
      Die von Kepler neu entwickelte dynamische Theorie fu{\ss}t auf der antiken Physik und der antiken Himmelsmechanik. Die antike Physik folgt der unmittelbaren Anschauung, dass n\"amlich alles Materielle aus den vier, bereits o.g. Elementen zusammengesetzt ist und dass jedem dieser Elemente offenbar eine Kraft ({\selectlanguage{greek}d\'unamis}) innewohnt, an den Ort zu gelangen, der ihm qua (g\"ottlicher) Ordnung ({\selectlanguage{greek}k\'osmos}) zugeteilt ist. Deswegen fallen Steine (auch im Wasser), Luftblasen steigen im Wasser auf und das Feuer leckt in die H\"ohe. Neben dieser \emph{vertikalen} Ordnungsbewegung -- aus der Eindeutigkeit des Ordnungsortes folgt insbesondere die Kugelgestalt der Erde mit genau einem Zentrum -- gibt es \emph{erzwungene} Bewegungen, die beseelte Wesen bewirken k\"onnen und es gibt die \emph{ewigen}, gleichm\"a{\ss}ig kreisf\"ormigen Bewegungen der Gestirne, das ist g\"ottliche Bewegung\footnote{\citet{prantl1857aristoteles-no-url}.}.
      
      Bei genauerer Betrachtung erwies sich etwa die Bewegung der Sonne -- sie l\"auft einmal im Jahr um die Erde entgegen der t\"aglichen Drehung des gesamten Fixsternhimmels -- als nicht gleichf\"ormig. Auf dem Kreis der \emph{Ekliptik}\footnote{wenn der Mond diesen Kreis schneidet, kann es zu Eklipsen, d.h. Finsternissen kommen, siehe Abschnitt \ref{scn:saros}.} -- auf ihm liegen ja auch die Tierkreiszeichen -- der sowohl zum Horizont als auch zum Himmels\"aquator geneigt ist, bewegt sich die Sonne im Winter schneller, im Sommer langsamer. Diese \emph{Anomalie} l\"asst sich durch einen Trick mit dem Axiom der ewigen,  gleichm\"a{\ss}ig kreisf\"ormigen Bewegung vers\"ohnen, indem man n\"amlich die Erde aus dem Drehzentrum herausr\"uckt (Exzenter), was, wie der Antike bereits bekannt war, gleichbedeutend damit ist, dass man die Sonne auf einem \emph{Epizykel} (Aufkreis) gleichm\"a{\ss}ig kreisen l\"asst, dessen Zentrum sich wiederum gleichm\"a{\ss}ig, in entgegengesetzter Richtung, mit gleicher Winkelgeschwindigkeit auf einem \emph{Deferenten} um die Erde dreht. 
      Noch genauer wurde man, wenn die Erde exzentrisch belassend eine weitere Aufweichung des o.g. Axioms zugelassen wurde, dass n\"amlich die Geschwindigkeit auf dem Deferenten konstant ist, nicht gesehen vom Zentrum des Deferenten, sondern von einem \emph{\"Aquanten}  (Ausgleichspunkt), der um eine Exzentrizit\"at vom Zentrum aus, entgegen der Richtung der aus dem Zentrum ger\"uckten Erde gelegt wird\footnote{\cite{van1960ausgleichspunkt}.}. 
      

      Mit \"ahnlichen Methoden lassen sich auch die Bewegungen der anderen \emph{Wandelsterne}, n\"amlich des Mondes und der Planeten, modellieren. Da die Planeten im Wesentlichen der Ekliptik folgen, hat ihr Deferent denselben Radius wie der des Sonnendeferenten. Der erste Epizykel modelliert die \emph{Scheinkreisbewegungen} gegen\"uber dem Sternenhintergrund. F\"ur dessen Periode ergibt sich f\"ur die \"au{\ss}eren Planeten bereits augenscheinlich als gute Sch\"atzung die Periode der Sonnenbahn um die Erde. Da sich aber das Zentrum des Epizykels auf dem Deferenten mit dem entsprechenden Bruchteil der \emph{siderischen Periode}\footnote{die erh\"alt man aus dem Fortschritt des Planeten gegen\"uber dem Sternenhintergrund innerhalb einer synodischen Periode (s.u.) unter Ber\"ucksichtung der dazu notwendigen Planeten- und Sonnenuml\"aufe. Zur besonderen astrologischen Bedeutung der Venus kam hinzu, dass ihre synodische Periode (ca. 584 Tage) zum Sonnenjahr (ca. 365 Tage) den Goldenen Schnitt approximiert, der im Rahmen des (pythagoreischen) Pentgramms im Abschnitt \ref{scn:pentagrm} angerissen wird.} weiterbewegt, weicht die \emph{synodische} Periode, nach der der Planet wieder denselben Stand bez\"uglich der Sonne einnimmt\footnote{z.B. Opposition f\"ur die \"au{\ss}eren Planeten und die erste Beobachtung als Abendstern f\"ur die inneren.}, von der Periode des Epizykels ab. Zweimal im Verlauf der synodischen Periode ist der Planet dann kurzzeitig \emph{station\"ar} (er steht scheinbar still).
      Wieder ist die Erde aus dem Drehzentrum jedes Deferenten herausger\"uckt. War die Exzentrizit\"at der Sonnenbahn nur etwa $3\,\%$\footnote{\citet[S. 58]{neugebauer1975history}.}, so betr\"agt sie $10\,\%$ f\"ur den Mars\footnote{\citet[S. 185]{neugebauer1975history}.}. Zu den Deferentbahnen kommen noch die Epizykelbahnen, die f\"ur Mars und Venus die Abst\"ande zur Erde auf 1/3 reduzieren bzw. auf 5/3 vergr\"o{\ss}ern k\"onnen\footnote{das ergibt sich aus den in der zweiten Spalte der obersten Tabelle von \citet[S. 185]{neugebauer1975history} angegebenen Epizykelradien zum Deferentenradius von 60.}.
      

      Bemerkenswerterweise ergibt sich an dieser Stelle die Dekonstruktion eines Mythos, 
      die in gewisser Analogie zur Dekonstruktion des Mythos steht, das Mittelalter habe an eine flache Erde geglaubt \citep{kruger2012moles}:
      
      \begin{quote}{\it 
      Im sogenannten \emph{geozentrischen Weltbild} steht die Erde geometrisch nur bez\"uglich des Fixsternhimmels im Zentrum, nicht aber bez\"uglich der Sonne und noch weniger bez\"uglich der Planetenbahnen\footnote{insbesondere m\"usste hier im Vorab der Begriff des Zentrums gekl\"art werden und bereits Aristoteles setzt sich kritisch mit der Uneindeutigkeit dieses Begriffs auseinander \citep[S. 157]{prantl1857aristoteles-no-url}. Heute k\"onnte man zudem fragen, handelt es sich um einen Median, n\"amlich einen Minimierer des \"uber Ort und Zeit gemittelten Abstands oder doch eher um einen Mittelwert, der \"uber die gemittelten Abstandsquadrate minimiert? Beide ergeben in der Regel unterschiedliche Orte. \label{foot:median-mean}}.}
      \end{quote}
      
      Mit \"Aquanten f\"ur jeden Deferenten (und ggf. weiteren verschachtelteten Epizyklen), auch f\"ur den Mond lie{\ss} sich eine noch h\"ohere Genauigkeit erreichen. 
      Ptolem\"aus stellte das bis ins zweite Jahrhundert gesammelte antike astronomische Wissen sorgf\"altig zusammen, korrigiert durch eigene Beobachtungen und Rechnungen und sein Lehrbuch der \emph{Almagest}\footnote{\cite{toomer1984ptolemy}.} war ma{\ss}geblicher Standard mit hoher numerischer Pr\"azision f\"ur das Mittelalter bis zur fr\"uhen Neuzeit\footnote{\citet[Band 1]{neugebauer1975history}.}. 
      
      \subsection{Weiterentwicklungen von Capella, Kopernikus und Brahe}
      
      Bereits in der Antike gab es Konkurrenzmodelle, die die Sonne in den Mittelpunkt des Kosmos setzten und die Erde und die Planeten um sie kreisen lie{\ss}en. Ein fr\"uher Verfechter war die Schule des Pythagoras ($\dagger$ ca. 495 v. Chr.), deren Ansichten Aristoteles diskutiert\footnote{\citet[S. 157]{prantl1857aristoteles-no-url}.}. Weitere Quellen, z.B. Archimedes ($\dagger$ ca. 212 v. Chr.),  Cicero ($\dagger$ 43 v. Chr.) und Plutarch ($\dagger$ ca. 125) verweisen auf Aristarch von Samos ($\dagger$ ca. 230 v. Chr.), der u.a. unter geschickter Ausnutzung des Strahlensatzes und genauer Beobachtung von Sonnen- und Mondfinsternissen auch die Radien von Erde, Sonne und Mond berechnete, sowie die Abst\"ande der beiden letzten von der Erde\footnote{z.B. \citet[S. 246]{heiberg1881archimedes}, \citet[S. 223]{heath1897works}, \citet[S. 630]{Cicero1933}, \citet[S. 139]{lesage2019critical}. Zu gro{\ss}en Teilen war er erstaunlich exakt, nur bez\"uglich Sonnenabstand und -durchmesser lag er etwa um den Faktor 20 zu niedrig, da ihm hinreichend exakte Winkelmessungen nicht m\"oglich waren.}. Diese Modelle erschienen allerdings sowohl naturwissenschaftlich fragw\"urdig (wie sollte die \"atherische Sonne die schwere Erde, auf die alles f\"allt, auf eine Bahn zwingen?) als auch gesellschaftlich-religi\"os (dass aus der Menge der G\"otter nur ein Sonnengott \"ubrig blieb, war eigentlich nur f\"ur Stoiker attraktiv)\footnote{das ergibt sich z.B. aus den von \citet{lesage2019critical} kritisch diskutierten Zitaten Plutarchs.}. Vermutlich aus diesem Grund sind entsprechende \"Uberlieferungen sp\"arlich.
      
      Dennoch war dem Mittelalter wohl bekannt, dass f\"ur die inneren Planeten das ptolem\"aische Modell (fast) \"aquivalent mit einem pythagoreischen ist\footnote{z.B. \citet{gingerich1985did}.}; diese  umrundeten ja auch offensichtlich nicht die Erde, sondern verblieben stets im \emph{Umkreis} der Sonne. 
      Visualisiert man die Beobachtungen von Sonne, Merkur und Venus im ptolem\"aischen Modell, so stimmen die Bahnen aller drei Deferenten nahezu vollst\"andig \"uberein (nach Konstruktion haben sie ja bereits identische Radien)\footnote{vgl. \citet[SS. 146 ff.]{neugebauer1975history}, vermutlich ist dieses Ph\"anomen bereits von Ptolem\"aus beabsichtigt \citep[z.B. Buch X, 4]{toomer1984ptolemy}.}. Somit beschreiben, grob gesagt, die beiden 
      gr\"o{\ss}ten Epizyklen von Merkur und Venus jeweils ihre Bahnen um die Sonne, die sich wiederum um die Erde dreht. Auch die Exzentrizit\"aten f\"ur die Bahnanomalien k\"onnen als Epizyklen \"ubernommen werden\footnote{verschobene Kreise bleiben Kreise.}. Da die 
      konstanten Winkelgeschwindigkeiten bez\"uglich der \"Aquanten nicht einfach \"ubernommen werden k\"onnen, ist das Modell numerisch etwas ungenauer, es sei denn, weitere Epizyklen w\"urden eingef\"ugt. 
      
      

   Ein solches Modell lernte Kopernikus ($\dagger$ 1543) im Studium als das Modell des Martianus Capella ($\dagger$ ca. 420)\footnote{\citet[S. 8, 2. Spalte]{copernicus1543revolutionibus}, siehe auch \citet[S. 221]{goldstein2002copernicus}.} kennen und offensichtlich versuchte er, dasselbe auch auf die \"au{\ss}eren Planeten anzuwenden. Eine naive Visualisierung der Beobachtungen von Sonne, Mars, Jupiter und Saturn im ptolem\"aischen Modell war dazu nicht geeignet, denn weder liefen die Deferenten ann\"ahernd \glqq im Takt\grqq, noch ergab es Sinn, dass etwa  Saturn, der langsamste aller Planeten, viel n\"aher an der Sonne st\"unde als etwa Merkur, der schnellste Planet. Hier stand die von Ptolem\"aus eingef\"uhrte Normierung im Weg, sein Deferent hatte immer gr\"o{\ss}eren Radius als der gr\"o{\ss}te Epizykel. Die grundlegende Leistung des Kopernikus bestand nun wohl darin, 
    f\"ur die \"au{\ss}eren Planeten Ptolem\"aus' Normierung  doppelt umzukehren: Zum Ersten jeweils den Deferent mit dem gr\"o{\ss}ten Epizykel zu vertauschen und zum Zweiten den neuen Deferent wieder auf den Radius des Sonnendeferenten zu normieren. Damit wurde der Radius des vormaligen Deferenten entsprechend gr\"o{\ss}er und in der Visualisierung ergab sich dasselbe Ergebnis\footnote{bereits in der Konstruktion der oben besprochenen ptolem\"aischen Epizyklen angelegt, vgl. auch \citet[S. 45]{toomer1984ptolemy}.}: Alle Deferenten drehten sich ann\"ahernd gleich mit der Sonne und somit auch die \"au{\ss}eren Planeten ann\"ahernd auf Bahnen um die Sonne. Mit 
    weiteren Epizyklen f\"ur die Planetenbahnanomalien war das im Wesentlichen das Modell des Tycho Brahe ($\dagger$ 1601), der zwar zeitlich nach Kopernikus kam, dessen letzten Schritt 
    aber nicht mitging\footnote{z.B. \citet{blair1990tycho}.}. 
      

      Dieser letzte Schritt bestand darin, den Koordinatenursprung von der Erde in die Sonne zu verlegen.
      Da die Reihenfolge in der Addition von Vektoren -- und damit die Reihenfolge der Addition von Kreisen -- vertauschbar ist, \"anderte sich nichts an der Planetenbewegung. 
      Das entsprechende \emph{kopernikanische Modell} ist somit fast vollst\"andig \"aquivalent mit dem ptolem\"aischen System, 
      allerdings komplexer\footnote{tats\"achlich nutzte  Ptolem\"aus zur h\"oheren Genauigkeit weitere kleinere Epizyklen \citep[z.B.: SS. 206 ff.]{neugebauer1975history}, Kopernikus brauchte indes mehr \citep{Gingerich1975CrisisVA}.}, wobei der jeweils erste Epizykel weitaus kleiner wurde. Da die  ptolem\"aischen Exzentrizit\"aten \"uber \"aquivalente und die \"Aquanten \"uber approximative Epizyklen dargestellt wurden, gilt nun auch:
      
      
      \begin{quote}{\it 
      Im sogenannten \emph{heliozentrischen Weltbild} steht die Sonne zwar definitionsgem\"a{\ss} im Drehzentrum aller Deferenten, faktisch (trickreicherweise) ist sie aber bis zu 10 $\%$ herausger\"uckt\footnote{
      Vermutlich nahmen einige Inquisitoren f\"alschlicherweise an, dass die Bahnanomalien  im ptolem\"aischen Modell nur durch Epizykel und nicht durch Exzenter und \"Aquanten modelliert w\"urden (vgl. auch Fu{\ss}note \ref{foot:median-mean}) -- bereits Nikolaus von Kues ($\dagger$ 1464) lehrte in seiner \emph{Docta ignorantia} (II, 12), dass das Zentrum der Welt nur theologisch gefasst werden k\"onne, physikalisch sei jeder Punkt ein Zentrum (\citet[S. 94]{kues1964diebelehrte}: Unde erit machina mundi quasi habens undique centrum et nullibi circumferentiam, quoniam eius circumferentia et centrum est Deus, qui est undique et nullibi)  --
      so dass sie zun\"achst 1616 und dann wieder 1633 von Galilei forderten, folgende S\"atze nicht zu lehren (so im vatikanischen Geheimarchiv, \cite{pagano2009documenti}): 
      \emph{Die Sonne ist das Zentrum der Welt und v\"ollig unbeweglich in \"ortlicher Bewegung} (Sol est centrum mundi et omnino immobilis motu locali) und \emph{die Erde ist nicht das Zentrum der Welt und nicht unbeweglich, sondern wird insgesamt bewegt, auch in t\"aglicher Bewegung} (terra non est centrum mundi, nec immobilis, sed secundum se totam movetur, etiam motu diurno). 
      Auch der hier offenbar als absolut vorausgesetzte  physikalische (und nicht theologische) Bewegungsbegriff war bereits von Nikolaus \citet[z.B. S. 94]{kues1964diebelehrte} relativiert worden, was sich schlie{\ss}lich  bei Newton ($\dagger$ 1727) (wohl \"uber Kepler) in seinem dritten Gesetz niederschlug \citep[Lex  III]{NewtonPrincipia}:
\emph{Zu jeder Wirkung gibt es immer eine entgegengesetzte und gleiche Gegenwirkung; oder die Wirkungen zweier K\"orper aufeinander sind immer gleich und sie richten sich nach entgegengesetzten Seiten aus} (actioni contrariam semper et aequalem esse reactionem: sive corporum duorum actiones in se mutuo semper esse aequales et in partes contrarias dirigi). 
      }.}
      \end{quote}
            
      \subsection{Keplers dynamische Theorie und Newtons Durchbruch}
      Tycho Brahe brachte verfeinerte astrometrische Methoden und damit exaktere Beobachtungsdaten aus D\"anemark nach Prag\footnote{\cite{dreyer1890tycho-notonline}.} und Kepler ($\dagger$ 1630), erst sein Assistent, dann sein Nachfolger folgte zwar der kopernikanischen Lesung des ptolem\"aischen Modells, st\"orte sich aber an dessen mangelnder Genauigkeit. Die \"Aquanten nicht verwerfend erkannte er sie offenbar irgendwann hellsichtig als Brennpunkte von Ellipsen\footnote{die ja mit einem Deferenten und einem Epizykel doppelter Frequenz exakt beschrieben werden.}. 
      In seiner \emph{Astronomia Nova} berichtet er von seinem 
      jahrelangen \glqq Krieg mit dem Mars\grqq~ und seines \glqq Sieges\grqq~\"uber ihn\footnote{\citet{kepler1609astronomia}, insbesondere zu Beginn in seiner bildreichen Widmung an Rudolf II.; eine Zusammenfassung dieses \glqq Krieges\grqq~(mit seinen Irrt\"umern entlang des Weges, die sich Kepler eingestand und wohl deshalb erfolgreich fortschreiten konnte) gibt \cite{gingerich1972johannes}.}, indem er zeigen konnte, dass sich dieser und alle anderen Planeten auf Ellipsen bewegten, in deren einem Brennpunkt die Sonne stand (Erstes Keplersches Gesetz). Die erstaunlich exakten Bahngeschwindigkeiten ergaben sich daraus, dass er den Inhalt des Fl\"achenst\"ucks, das vom Strahl von der Sonne zum Planeten in gleicher Zeit \"uberdeckt wurde, konstant setzte (Zweites Keplersches Gesetz)\footnote{\citet{wilson1972did}, das entspricht dem modernen Impulserhaltungssatz.}. Dieses Modell war ungleich einfacher, da es vollst\"andig ohne Epizyklen auskam und es war exakter als das ptolem\"aische Modell. Zwar war damit das antike Prinzip von der ewigen, gleichf\"ormig kreisf\"ormigen Bewegung zu Grabe getragen\footnote{indes konnte Feynman \citep{goodstein1996feynman-notonline} zeigen, dass sich Keplers Ellipsen \"uber geeignete Projektionen einer gleichf\"ormigen Kreisbewegung erhalten lassen.}, aber es  erlaubte ihm, die Planetenbewegungen recht genau im Jahr 7 vor Christus zu rekonstruieren\footnote{\cite{kepler1614DeVeroAnno}.}. 
      
      Bislang waren die kosmologischen Modelle nur 
      mechanischer Natur\footnote{Galileis Versuch \"uber seine Erkl\"arung von Ebbe und Flut  das ptolem\"aische System zu falsifizieren, \"uberzeugte die zeitgen\"ossische \glqq wissenschaftliche Community\grqq~nicht \citep[S. 40-44, 274]{drake2003galileo}.
      Korrekterweise glaubte Galilei allerdings, dass die Erddrehung eine anderweitig gleichf\"ormige Bewegung ablenken w\"urde, dieser Effekt sei aber zu klein, um gemessen zu werden \citep[S. 279]{galilei1632dialogo}. Diese  \emph{Scheinkraft}, seit 1835 bekannt als \emph{Corioliskraft}, wurde 1851 mittels des nach Foucault benannten Pendels nachgewiesen. Bemerkenswerterweise w\"are aber selbst mit diesem Nachweis das ptolem\"aische System nicht falsifiziert, denn es war nur ein mechanisches System. W\"are es um eine dynamische Theorie erg\"anzt worden --  Kepler entwickelte seine eigene dynamische Theorie (s.u.), die Galilei aber wohl abhlehnte \citep[S. 40]{drake2003galileo} -- dann h\"atte die Erg\"anzung diese Scheinkraft aufnehmen m\"ussen und w\"are damit weiterhin g\"ultig.
      }. 
      Wenn nun die Sonne die schwere Erde auf eine Bahn um sie zwang, dann musste sie nach der weiterhin g\"ultigen antiken Physik beseelt sein\footnote{so dachte auch Ptolem\"aus, wie \citet[SS. 38/39]{murschel1995structure} erkl\"art.}. Seltsamerweise geschah die Kraft\"ubertragung ohne Ber\"uhrung\footnote{ein Unm\"oglichkeit, wie Thomas v. Aquin in seiner Summa contra gentiles (ScG 2,20,6) lehrt, vgl. auch \citet[S. 268]{kochiras2009gravity}.}. Als Erkl\"arung bot sich nur das bislang mit Seelenkraft erkl\"arte Ph\"anomen des Magnetismus an\footnote{\citet{sander2018magnetism}.}. Damit blieben die Himmelsk\"orper beseelt und konnten auf die Seelen der Menschen einwirken: Die Grundlagen der Astrologie blieben erhalten. 
      
      Unabh\"angig von seinem empirisch gewonnenen Dritten Gesetz ($T^2/R^3$ ist eine universale Konstante f\"ur alle Planeten, wobei  $R$ die jeweilige Bahnhalbachse  und $T$ die entsprechende Umlaufzeit bezeichnet) ging Kepler davon aus,  dass die Sonnenkraft mit dem inversen Abstand abn\"ahme, so als ob die Kraft auf einem Kreis verteilt w\"are, dessen Umfang proportional zu ihrem Radius ist\footnote{\citet[Cap. 33, S. 170]{kepler1609astronomia}.}. Tats\"achlich entspricht sein Drittes Gesetz aber einer Abnahme der Sonnenkraft mit dem inversen Abstandsquadrat, die Kraft ist also auf einer Sph\"are verteilt, deren Oberfl\"ache proportional zum Quadrat ihres Radius ist\footnote{In erster N\"aherung ist mit der Geschwindigkeit $v$ die Umlaufzeit gegeben durch $T=2\pi R/v$ und die Beschleunigung durch $a = v^2/R = 4\pi^2 R/T^2 = c/R^2$ unter der Annahme, dass $4\pi^2 R^3/T^2$ eine universelle Konstante $c$ ist, vgl. \citet[S. 245]{gal2019between}}.

      Darauf aufbauend stellte Newton ($\dagger$ 1727) seine allgemeine Theorie der Gravitation auf und nach mehrj\"ahriger M\"uhe\footnote{\cite{cohen1981newton}.} konnte er zeigen, dass die drei Keplerschen Gesetze aus seiner Differentialgleichung folgten\footnote{vgl. \citet[I, Prop XI]{NewtonPrincipia},  eine Punktmasse $M$ am Koordinatenursprung beeinflusst nach \citet[III, Prob. VII und VIII]{NewtonPrincipia}, modern ausgesagt, den Pfad $\vec r = \vec r(t)$ eines Teilchens  zum Zeitpunkt $t$ in Entfernung $r=r(t)$ vom Ursprung durch $\frac{d\vec v}{dt} = \ddot{\vec r} = -\frac{kM}{r^2} \frac{\vec r}{r} $, wobei $k> 0$ konstant ist. 
      }. Offen blieb und bleibt bis heute, wie die Kraft tats\"achlich \"ubertragen wird\footnote{Naturwissenschaft leistet nicht die Eliminierung, sondern ein Verschieben des Geheimnisses.}.

   \section{Hintergr\"unde der Magier: Das gro{\ss}e Forschungsprojekt}
   
   Kehren wir zur\"uck in die Zeit um Christi Geburt und zum damaligen Hintergrund von Gelehrten (Magiern) aus dem Osten. Seit etwa dreieinhalbtausend Jahren hatten sich die V\"olker im Zweistromland, so wie es Ausgrabungen von Keilschrifttafeln\footnote{\cite{sharlach2013calendars}.},  u.a. sogenannten \emph{Omina}, 
   zeigen, mit der Beobachtung von Himmelsereignissen, von Ereignissen auf der Erde, ihrer Zuordnung und Vorhersage besch\"aftigt\footnote{\cite{ossendrijver2011himmel}, eine gr\"o{\ss}ere Auswahl bringt \cite{gossmann1950planetarium}.}. Modern gesprochen konnten im Zuge dieses \"au{\ss}erst langfristig angelegten Forschungsprojekts u.a. Perioden von Himmelsereignissen mit gro{\ss}er Pr\"azision gefunden werden und dazu und zur Astrologie (Korrelation von Himmelsereignissen mit irdischen Ereignissen) wurde \glqq Big Data\grqq~produziert, eine schier un\"ubersichtliche Anzahl von Omina, von denen vermutlich nur ein Bruchteil bislang geborgen ist. Um Sonnen-,  Mond- und Sternperioden, die nicht in ganzzahligen Verh\"altnissen zueinander stehen, doch miteinander in Bezug zu setzten, entwickelten die Babylonier ein \emph{innerliches} (esoterisches) Wissen, mit dem sie nicht nur einen stabilen Kalender schufen, sie waren auch in der Lage, Mond- und Sonnenfinsternisse vorherzusagen\footnote{sehr exakt f\"ur Mondfinsternisse, eher approximativ f\"ur Sonnenfinsternisse, vgl. Abschnitt \ref{scn:saros}.}. Das sicherte ihren Gelehrten Einfluss und Auskommen, modern gesprochen einen stetigen Flu{\ss} von Forschungsmitteln. Offenbar aus gutem Grund \"uberlieferten sie nur ihre Ergebnisse, nicht aber ihre Methoden, die aus heutiger Sicht allerdings zumeist gut rekonstruierbar erscheinen\footnote{\citet[Buch 2]{neugebauer1975history}.}. 
   
   
   \subsection{Esoterik f\"ur Anf\"anger: Kalender und das kleinste gemeinsame Vielfache}
   Eine \emph{Lunation} (von einem Neumond zum n\"achsten) dauert ca. $29,5306$ Tage\footnote{das entspricht $29$ Tagen, $12$ Stunden und etwa $45$ Minuten. Die Babylonier hatten ein Zahlensystem, das auf der Zahl $60$ aufbaute und mit drei Stellen hinter dem Komma ganzzahlige Vielfache von $1/60, 1/3600$ und $1/216000$ darstellen konnte, vgl. \citet[Band 2]{neugebauer1975history}.}, somit sind nach einem \emph{Mondjahr} mit $6$ \emph{hohlen Monaten} zu $29$ Tagen und $6$ \emph{vollen Monaten} zu $30$ Tagen insgesamt $354$ Tage vergangen. Von einer Tag-und-Nachtgleiche zur \"ubern\"achsten sind es ca. $365,25$ Tage, dieser Zeitraum wird nach der Kalenderreform durch Julius Caesar (zum 1. Jan. 45 v. Chr.) als \emph{julianisches Jahr} bezeichnet\footnote{einen weiten \"Uberblick \"uber verschiedene historische Kalendersysteme gibt \citet{stern2012calendars}.}, das damit $11,25$ Tage l\"anger ist als ein Mondjahr. Um dennoch Mond- und Sonnenjahr zu synchronisieren, lautete die Methode der Babylonier:
   \begin{center} {\it
    Finde das kleinste gemeinsame Vielfache (kgV).}
   \end{center}
   Wie man leicht nachrechnet, sind nach $125$ vollen Monaten und $110$ hohlen Monaten genau $19$ julianische Jahre und ein Viertel-Tag vergangen, selbst die Synchronisation mit der tats\"achlichen Mondphase hat sich nur um ca. einen Drittel-Tag verschoben\footnote{$ 125 \cdot 30 + 110 \cdot 29 - 235 \cdot 29,5306  =  0,309$.}. In dem nach Meton (aus dem 5. Jhd. v. Chr.) benannten Kalender\footnote{z.B. \citet{cohen2016problem}.} werden somit im Verlauf von 19 Mondjahren 7 volle \emph{Schaltmonate} eingef\"ugt und gleichzeitig wurden 4 hohle Monate zu vollen\footnote{etwas einfacher ist es, ein \glqq ganzzahliges\grqq~Sonnenjahr mit nur 365 Tagen mit einem um 11 Tage kürzeren Mondjahr von 354 Tagen zu harmonisieren: Man f\"uge innerhalb von 19 Jahren 6 volle Monate und einen hohlen Monat ein, denn es gilt:  19 ·  11 = 190 + 19 = 180 + 29 =  6 · 30 + 1 · 29. Noch genauer ist die \emph{Methode der \"Agypter}, siehe Fu{\ss}note \ref{foot:aegypt}.}. 
   Eine einfache Regel f\"ur die Notwendigkeit, Schaltmonate einzuf\"ugen, war die Beobachtung des 
   \emph{Neulichts}, die erste Mondsichel nach Neumond, sichtbar am Abendhimmel, direkt neben der Sonne.  Wenn diese im Stier, genau zwischen Plejaden und Hyaden stand, die beiden Sternhaufen bildeten das \emph{Goldene Tor der Ekliptik}, das die Planetenbewegungen und im Wesentlichen auch die des Mondes begrenzte\footnote{
   daraus ergibt sich z.B. die \emph{Plejadenschaltregel}:
   \emph{Wenn am 1. Nisan Plejaden und Mond in Konjunktion stehen, so ist dieses Jahr normal; wenn erst am 3. Nisan, so ist dieses Jahr ein Schaltjahr}, so \citet[S. 109 u. 186]{gossmann1950planetarium}, vgl. auch \citet[S. 3, der offenbar eine falsche Seite bei G\"ossmann angibt]{miller1988pleiades}. 
   In der Tat steht der Mond nach seiner \emph{siderischen} (bez\"uglich des Sternenhintergrunds) Periode von ca. 27,3217 Tagen definitionsgem\"a{\ss} an genau derselben Stelle des Sternhintergrunds, aber an einem vollen Phasendurchgang fehlen ca. $29,5306 -27,3217 = 2,2089$ Tage. 
   \label{foot:siderischerMond}
   Nach einem Mondjahr geht der Mond siderisch um etwas mehr als einem Tag nach, denn er steht erst nach ca. $13\cdot 27,3217 = 354 + 1,1821$ Tagen wieder an derselben Stelle vor dem Sternenhintergrund. Mit  einem  vollen Schaltmonat l\"asst sich nun der Mond  siderisch um ca. $30 - 27,3217 = 2,6783$ Tage vorsetzen, mit einem hohlen um ca. $1,6783$ Tage.
    Nach ausf\"uhrlicher Diskussion bei \cite{papke1984zwei} findet sich diese 
    Regel bereits vor mehr als dreitausend Jahren auf babylonischen Tafeln.
    
    Zur Illustration ihrer hohen Praktikabilit\"at wird diese Regel kurz angespielt: 
    Nach einem lunaren Jahr und einem Tag hat der Mond, wie bereits oben erw\"ahnt,  in etwa dieselbe Position gegen\"uber 
    dem Fixsternhimmel. Stand er also in einem ersten Jahr zum 1. Nisan im Goldenen Tor, so steht er dort im dritten lunaren Jahr erst zum 3. Nisan. Wird dann dieser angefangene Nisan zu einem vollen Schaltmonat und folgt erst danach der Nisan, so hat sich zu diesem neuen 1. Nisan im Vergleich zum Fixsternhimmel des ersten Jahres zum 1. Nisan die Sonne ca. $360 * (30 - 2*11,25)/365,25 \approx 8$ Grad \glqq zu weit\grqq~nach Osten bewegt, d.h. im folgenden (vierten) lunaren Jahr steht sie zum 1. Nisan fast wieder an derselben Position und der Mond fast wieder im Goldenen Tor (genauer erst nach $24*(40*27,3217 - 3*354 - 30) \approx 21$ Stunden). Dann steht im 7. lunaren Jahr zum 3. Nisan wieder der Mond im Goldenen Tor und ein neuer Schaltmonat wird vor diesen Nisan gesetzt...
    
    Hier stellen \cite{hansen2008himmelsscheibe} einen Bezug zur \emph{Himmelscheibe von Nebra} her, die demnach eine Anleitung zur Kalenderkorrektur durch Schaltmonate darstellt. Da diese \emph{Plejadenschaltregel} aufgrund der Pr\"azession der Erdachse am 9.4.631 spektakul\"ar versagt haben soll, schlie{\ss}en \cite{rink2013deraltarabische} deshalb auf eine Kalenderreform der Araber, die auch von einer Inschrift Muawijas, datiert ca. 663, als \emph{Zeitrechnung der Araber} (\cite{hirschfeld1981roman}, diese Inschrift beginnt mit einem Kreuz obschon Muawija in der islamischen Chronologie als der f\"unfte Kalif gilt) best\"atigt wird.
    },
   dann koinzidierte dieses Ereignis vor gut 4.000 Jahren einmal im Verlauf eines Metonzyklus (d.h. alle 19 Jahre) mit dem Gang der Sonne durch die Fr\"uhlingstag-und-Nachtgleiche%
    \footnote{dieser Punkt ist ca. 60 Grad von der heutigen Fr\"uhlingstag-und-Nachtgleiche entfernt und ergab also die Fr\"uhlingstag-und-Nachtgleiche vor ca. 26.000/6 = 4.333 Jahren, vgl. die Bemerkung zur Pr\"azession der Erdachse zu Ende dieses Abschnitts.
   Sinnf\"allig w\"are eine entsprechend adaptierte Plejadenschaltregel auch um den Herbstanfang vor ca. 4.000 Jahren, da im Verlauf eines Metonzyklus ein (ziemlich exakter) Vollmond genau einmal zwischen Plejaden und Hyaden beobachtbar w\"are -- und das die ganze Nacht lang. Vielleicht begannen die r\"omischen Steuerjahre (\emph{indictiones}), die wohl auf dem antiken \"agyptischen Kalender beruhen \citep[S. 734-735, 739]{fotheringham1929calendar}, aus diesem Grund um die Herbsttag-und-Nachtgleiche? Um diesem Zeitpunkt lag damals allerdings auch der heliakalische Aufgang des Sirius, dessen Beobachtung den Beginn des \"agyptischen Jahres begr\"undete \citep[S. 599 ff.]{neugebauer1975history}. Die \"Agypter synchronisierten ihr Sonnenjahr mit dem Mondjahr \"uber einen Zyklus von 25 Jahren (erkl\"art in Fu{\ss}note \ref{foot:aegypt}), der r\"omische Steuerzyklus dauerte allerdings nur 15 Jahre \citep[S. 739]{fotheringham1929calendar}. 
   }%
   , das konnte als Eichung verwendet werden\footnote{exakt war diese allerdings nur, wenn Neulicht eine der m\"oglichen 19 Mondphasen im Metonzyklus zur Tag-und-Nachtgleiche war und diese 19 Phasen \"anderten sich ca. alle 230 Jahre um einen Tag.}. %

   In der o.g. Kalenderreform durch Caesar wurden dann die Monate auf ihre heutige Gestalt so verl\"angert, dass deren $12$ ein Jahr von $365$ Tagen ergeben und jedes vierte Jahr durch einen Schalttag verl\"angert wurde. Der Meton-Zyklus angewandt auf diesen \emph{julianischen Kalender} lag vermutlich der Festlegung des Osterfests auf den ersten Sonntag nach dem ersten Vollmond nach der Fr\"uhlingstag-und-Nachtgleiche vom Konzil von Niz\"aa (325) zugrunde\footnote{z.B. \citet{cohen2016problem}.
   Der Fr\"uhlingsbeginn ist auf den 21. M\"arz festgelegt und nicht auf das tats\"achliche Passieren der Tag-und-Nachtgleiche durch die Sonne. Im Jahr 2019 geschah dieses am Abend des 20.3. mitteleurop\"aischer Zeit (kurz vor Mitternacht Jerusalemer Zeit) und so wurde der Vollmond kurz nach Mitternacht am 21.3. nicht gez\"ahlt: Ostern folgte erst einen Monat sp\"ater.
   }: 
   In einem 19-j\"ahrigen Metonzyklus gibt es ja 19 M\"oglichkeiten f\"ur einen Vollmond nach der Fr\"uhlingstag-und-Nachtgleiche. Unter Einbeziehung m\"oglicher Schaltjahre gibt es im julianischen Kalender damit einen Zyklus von $4 \cdot 19 = 76$ Kalenderdaten, auf die der erste Fr\"uhlingsvollmond fallen kann. Da es 7 weitere m\"ogliche Daten f\"ur den n\"achsten Sonntag gibt, ergibt sich ein Zyklus von $7 \cdot 76 = 532$ Jahren f\"ur die Wiederkehr desselben Kalenderdatums f\"ur das Osterfest. Dieser Zyklus wird weiterhin in der Ostkirche genutzt.
   
   Da sich im julianischen Kalender alle 100 Jahre die Tag-und-Nachtgleichen um knapp einen Tag verschieben\footnote{genauer dauert es 365,2422 Tage von einer Tag-und-Nachtgleiche zur \"ubern\"achsten.}, hinkte der julianische Kalender dem wahren Sonnenstand zu Keplers Zeiten bereits etwa 10 Tage hinterher.  Die Kalenderreform (1582) von Papst Gregor XIII. lie{\ss} also 10 Tage ausfallen und verkomplizierte die Osterfestberechnung der Westkirche\footnote{\citet[S. 355]{reich1990problems}.}. Innerhalb eines Jahrhunderts\footnote{im gregorianischen Kalender entfallen die Schaltjahre in jedem durch 100 teilbaren Jahr, nicht aber in den durch 400 teilbaren.} kann aber der entsprechend verschobene Metonzyklus der Ostkirche genutzt werden. Mittlerweile unterscheidet sich der julianische Kalender um 13 Tage vom gregorianischen\footnote{
   Obschon 
   laut \url{https://www.timeanddate.com/calendar/seasons.html?year=300} (6.3.2026) die Sonne die Tag-und-Nachtgleiche am Morgen des 21. M\"arz 325 passierte, so geschah dieses im 4. Jhd. mehrheitlich bereits am 20. M\"arz und somit passiert sie diese im fortlaufenden julianischen Kalender seit dem 20. Jhd. bereits zum 8. M\"arz (julianisch), auf die aber kalendarisch dann noch 13 Tage gewartet werden muss. Deswegen f\"allt bis zum Ende des 21. Jhd. der Weihnachtstermin der Ostkirche auf das westliche Dreik\"onigsfest, ab dem 22. Jhd. liegt er dann einen Tag sp\"ater. Die Abweichung des sp\"atantiken (den nutzt die Ostkirche) vom tats\"achlichen Metonzyklus geht ca. alle 230 Jahre um einen Tag in die andere (!) Richtung. Stimmten sie um die Zeitenwende miteinander \"uberein, so differieren sie aktuell um 9 Tage. F\"ur die Osterfestberechnung bis zum Ende des 21. Jhd. ergibt sich also folgende Faustregel: Ist der erste Fr\"uhlingsvollmond im gregorianischen Kalender nach dem 30.3., so stimmen West- und Ostostern \"uberein, falls der Vollmond auf einen Sonntag oder Montag fiel (zum Ende der Osternacht steht dann der Halbmond im S\"uden) -- das passiert ca. viermal (ca. 2/3 * 2/7) alle 19 Jahre. Sonst ist Ostostern eine Woche sp\"ater. War der erste Fr\"uhlingsvollmond vor dem 30.3., so liegt Ostostern einen Monat nach Westostern, bzw. einen Monat und eine Woche.
   }. 
   
   Bez\"uglich der Astrologie kommt ein weiterer Aspekt zum Tragen: Die \emph{Pr\"azession} der Erdachse, die vermutlich bereits antike \"agyptische Gelehrte beobachtet hatten\footnote{\citet{quack2018astronomy}, deren Messung erfolgte wohl erst durch Hipparchos (ca. 190--120 v. Chr.), vgl. \citet[S. 292 ff.]{neugebauer1975history}, den Babyloniern war sie wohl unbekannt, siehe \citet[S. 543]{neugebauer1975history}.}, verschiebt mit einem Zyklus von etwa $26$ Tausend Jahren den \emph{Fr\"uhlingspunkt} und damit alle Tierkreiszeichen in ca. $2166$ Jahren um ein Sternbild: Damit ist z.B. eine heute im astrologischen Sternbild der Jungfrau Geborene tats\"achlich ein \emph{L\"owe}.  
     
   \subsection{Esoterik f\"ur Fortgeschrittene: Saroszyklen der Finsternisse}\label{scn:saros}
    Eine Sonnen- oder Mondfinsternis kann nur bei Neu- bzw. Vollmond eintreten, hierf\"ur gilt die o.g. Lunation von ca. $29,5306$ Tagen. Zudem ist es notwendig, dass der Mond fast genau in der Ekliptik steht. Ein \emph{drakonitischer Monat}, das ist die Zeit zwischen zwei Ekliptikdurchg\"angen des Mondes in gleicher Richtung, dauert ca. $27,2122$ Tage\footnote{
    \label{foot:mondrotation}
    da sich die Rotationsebene des Mondes von der Erde aus gesehen mit einer Periode von ca. 18,6 Jahren dreht (z.B. \cite{meyer2011precession}), ist der drakonitische Monat entsprechend k\"urzer als der siderische (vgl. Fu{\ss}note \ref{foot:siderischerMond}).}. Das kgV ergibt nun eine Periode f\"ur Sonnen- bzw. Mondfinsternisse. Wie man wieder leicht nachrechnet, entsprechen 
    $242$ drakonitische Monate $18$ julianischen Jahren, $10.5$ Tagen (ein halber Tag fehlte an den 18 julianischen Jahren, um wieder auf ganze Tage zu kommen), $8$ Stunden und ca. 27 Minuten\footnote{$27,2122 \cdot 242 = 6585,3524 = 18\cdot 365,25+10,5 +0,3524$}. 
    $223$ Lunationen entsprechen $18$ julianischen Jahren, $10.5$ Tagen, $7$ Stunden und ca. 46 Minuten\footnote{$223 \cdot 29,5306 = 6585,3238$. 
    }. 
    Dieser letzte Zeitraum wird \emph{Saroszyklus} genannt\footnote{\citet[S. 502 ff]{neugebauer1975history}.}. Da die jeweilige Finsternis damit insbesondere um ca. 8 Stunden, d.h. um etwa einen Drittel-Tag versetzt stattfindet, ist sie auf der Erde etwa 120 L\"angengrade nach Westen versetzt beobachtbar und kehrt erst im dritten Zyklus zur\"uck, allerdings versetzt um einige wenige L\"angengrade und Breitengrade\footnote{
    der Unterschied zwischen drakonitischer und siderischer Periode (vgl. Fu{\ss}noten \ref{foot:siderischerMond} und \ref{foot:mondrotation}) ist verantwortlich f\"ur die Ver\"anderung der Breitengrade (es gibt ansteigende und absteigende Zyklen): Ein Saroszyklus entspricht 241 siderischen Monaten und etwa 20 Stunden. Damit ist der Mond nach einem Saroszyklus vor dem Sternenhimmel etwa $10^\circ$ nach Osten gewandert, und einige Grade nach Norden oder S\"uden. Die ver\"anderte H\"ohe bewirkt den Unterschied der Breitengrade. Die u.a. in S\"uddeutschland am 11.8.1999 beobachtete totale Sonnenfinsternis (Saroszyklus 145, vgl. \url{https://de.wikipedia.org/wiki/Saroszyklus}) war als totale Sonnenfinsternis am 21.8.2017 in den USA beobachtbar, sie wird am 2.9.2035 vom Nordchina bis in den Pazifik sichtbar sein und kehrt in unsere L\"angengrade am 12.9.2053 zur\"uck, beobachtbar als totale Sonnenfinsternis u.a. an der Mittelmeerk\"uste Nordafrikas. Saros 145 war bereits verantwortlich f\"ur 22 Finsternisse und es werden insgesamt 77 sein (\url{https://eclipse.gsfc.nasa.gov/SEsaros/SEsaros.html}).}. Aufgrund jahrtausendelanger Aufzeichnungen, waren diese langfristigen Zusammenh\"ange Gelehrten in der Antike bekannt\footnote{\citet[S. 502 ff]{neugebauer1975history}.}. Insbesondere sind in jedem Jahr etwa zwei Saroszyklen aktiv, damit ist  an einem festen Punkt auf der Erde pro Jahr im Schnitt etwa 
    eine (ggf. partielle Mond-) Finsternis vorhersag- und bei gutem Wetter beobachtbar.

   \subsection{Esoterik f\"ur Profis und antike Kritik: Der Kosmos ist nicht kosmisch}\label{scn:pentagrm}
   
   Obschon die Methode des kgV \"au{\ss}erst erfolgreich war, mussten mit genauerer Beobachtung von Himmelsereignissen immer gr\"o{\ss}ere Vielfache genommen werden. So kam bereits in der  Antike die Frage auf, ob die G\"otter die Welt so eingerichtet h\"atten, ob alle  Verh\"altnisse letztlich doch Br\"uche ganzer Zahlen seien, also Harmonien seien\footnote{auch Kepler versuchte die hinter den Bahnverh\"altnissen der Planeten liegenden, vermutet g\"ottlich gesetzten Akkorde zu entschl\"usseln, z.B.: \citet[Kap. 7, Buch V,  S. 212]{kepler1619harmonices}.}? Wenn ja, dann m\"ussten sich doch bei gro{\ss}em Rechenaufwand 
   schlie{\ss}lich die exakten Verh\"altnisse etwa des Jahres zum Monat und zum Tag finden lassen\footnote{\label{foot:aegypt} der \"agyptische Sonnenkalender hatte 365 Tage. Somit entsprachen einem  Zyklus von 25 \"agyptischen Sonnenjahren fast genau 309 Lunationen: $365\cdot 25 - 309\cdot 29.5309 = 0.0481$, d.h. erst nach 520 Jahren betrug der Unterschied in der Mondphase mehr als einen Tag \citep[S. 563]{neugebauer1975history}. Von den \"Agyptern \"ubernahmen wohl die R\"omer auch -- und von diesen wir -- die Einteilung des Tages in 24 Stunden; somit konnten sie den Tagesbeginn der \"Agypter zu Sonnenaufgang auf Mitternacht verlegen \citep[S. 560 ff.]{neugebauer1975history}.}. Letztlich lehrten u.a. das die Pythagoreer, aber bereits in ihrem Symbol, dem Pentagramm\footnote{zun\"achst wohl, weil es symmetrisch und hinreichend komplex war und doch in einem Zug ohne Absetzen gezeichnet werden konnte \citep[S. 149]{ball1892mathematical}.} stehen die Seitenverh\"altnisse im \emph{Goldenen Schnitt} zueinander  und dieses Verh\"altnis ist kein Bruch zweier ganzer Zahlen (rational), sondern irrational. F\"ur jede Seite gilt n\"amlich, dass sich ihre Gesamtl\"ange zur L\"ange des gr\"o{\ss}eren Teils verh\"alt wie die L\"ange des gr\"o{\ss}eren Teils zu der des kleineren\footnote{Bezeichnet man die Spitzen des Pentagramms im Uhrzeigersinn mit A,B,C,D,E und die jeweils gegen\"uberliegenden Ecken des inneren F\"unfecks mit a,b,c,d,e, so sind etwa die Dreiecke EAc und DBC \"ahnlich. Folglich gilt $\frac{\overline{BD}}{\overline{EA}}=\frac{\overline{BC}}{\overline{Ac}}$. Da  $\overline{EA}=\overline{BC}$, $\overline{AD}=\overline{BD}$ und DcBC eine Raute darstellt, folgt in der Tat $\frac{\overline{AD}}{\overline{cD}}=\frac{\overline{cD}}{\overline{Ac}}$.}. Durch L\"osen der sich so ergebenden quadratischen Gleichung\footnote{mit $0<1-x< x<1$ gelte $\frac{1}{x} = \frac{x}{1-x}$. Dann folgt $x = \frac{\sqrt{5}-1}{2} = \frac{\sqrt{5}-1}{2} 
   \cdot \frac{\sqrt{5}+1}{\sqrt{5}+1} = \frac{2}{\sqrt{5}+1}$.}   erh\"alt man das Verh\"altnis des Goldenen Schnitts als
   $$ \frac{\sqrt{5}+1}{2}\,.$$
   Nun l\"a{\ss}t Platon ($\dagger$ ca. 348 v. Chr.) im Dialog Theaitetos vom gleichnamigen J\"ungling den Mythos, $\sqrt{5}$ sei ein Bruch zweier ganzer Zahlen, dekonstruieren\footnote{147d-148b, eine sch\"one Erkl\"arung liefert \cite{ofman2014understanding}.}. Steht damit nicht zu bef\"urchten, dass es entweder keine g\"ottliche Kraft gibt, oder dass, wenn sie existiert, sie die verschiedenen kosmischen Gr\"o{\ss}en ebenso inkommensurabel eingerichtet hat, sie letztlich also keine Harmonie, die durch Verh\"altnisse ganzzahliger Br\"uche bestimmt ist, dem Menschen erlaubt? W\"are dann der Kosmos weniger kosmisch, d.h. weniger sch\"on und geordnet?    
   

   \subsection{Abgleich mit dem babylonischen Archiv}\label{scn:omina}
   Aufgrund der umfangreichen Forschungst\"atigkeit der Babylonier, ihrer guten Dokumentation und der Fortschritte der Arch\"aologie sind wir ohne Weiteres in der Lage, aus den Archiven der Babylonier ihre astronomischen Beobachtungen um das Jahr 7 v. Chr. zu entnehmen. Die Lage ist sogar so komfortabel, dass mindestens 3 nahezu identische, allerdings heute nicht mehr vollst\"andig erhaltene Keilschriftkopien\footnote{im British Museum unter BM 35429, BM 34659, BM 34614.} erstellt wurden\footnote{\cite{sachs1984kepler}.}. F\"ur Mars, Jupiter und Saturn l\"asst sich entnehmen\footnote{ebd. S. 47 ff., unter der Hypothese der Korrektheit der Aufzeichnungen der Sommersonnenwende am 29. Tag des Monats III (20., 21., oder 22. Juni) und der Wintersonnenwende am 5. Tag des Monats X (21. oder 22. Dezember) und der Angabe der jeweiligen Monatsl\"angen (alternierend zwischen 29 und 30 Tagen), lassen sich die o.g. Daten (sogar eindeutig) in unseren Kalender konvertieren (w\"ahlt man das letztm\"ogliche Datum, 22.6. f\"ur III 29, so entspricht X 5 dem fr\"uhstm\"oglichen Datum, 21.12), die entsprechenden Daten sind in Klammern angegeben, wobei negative Jahreszahlen das entsprechende Jahr v. Chr. bezeichnen.}: 
   {\it \begin{description}
    \item[Monat I, Tag 3:] Jupiter in den Fischen, erste Sichtbarkeit Saturns in den Fischen (30.3.-7), 
    \item[Monat IV, Tag 22:] Jupiter station\"ar (13.7.-7), 
    \item[Monat IV, Tag 29:] Saturn station\"ar (20.7.-7), 
    \item[Monat VI, Tag 21:] Jupiter und Saturn in Opposition (10.9.-7),
    \item[Monat VIII, Tag 19:]\footnote{die erste Ziffer des Datums ist unklar, es muss sich aber um 19 handeln, da auch anderenorts (z.B. \citet[Table 7]{de2019study}) Zeiten von ca. 120 Tagen (das entspricht modernen Angaben) f\"ur die Dauer der retrograden Bewegung (von der ersten Station zur zweiten) des Jupiters angegeben sind. Dagegen erscheint die Dauer der retrograden Bewegungen von Saturn hier, zwar in ungef\"ahrer \"Ubereinstimmung etwa mit \citet[Table 1, mit z.B. 118 Tagen]{steele2019early}, zu klein gegen\"uber modernen Angaben (ca. 4,5 Monate). F\"ur moderne Angaben siehe z.B. \cite{keller2021kosmos}.} Jupiter station\"ar (6.11.-7), 
    \item[Monat VIII, Tag 21:] Saturn station\"ar (8.11.-7), 
    \item[Monat X, Tag 26:] Mars erreicht die Fische (11.1.-6),
    \item[Monat XI, Tag 22:]\footnote{eine Variante gibt Tag 21.} Jupiter verl\"asst die Fische und geht in den Widder (5.2.-6),
    \item[Monat XII, Tag 12:]\footnote{eine Variante gibt Tag 9.} Saturn zum letzten Mal in den Fischen sichtbar (25.2.-6).
   \end{description}
   } 
   \noindent
   Bemerkenswerterweise enthalten diese astronomischen Aufzeichnungen weder irdische Ereignisse, geschweige denn eine astrologische Zuordnung. Hatten die Babylonier gegen Ende ihres Forschungsprojekts die Astrologie aufgegeben? Auch f\"allt auf, dass die dreifache Gro{\ss}e Konjunktion sich zwar aus den Beobachtungen ablesen l\"asst, sie aber nicht eigens erw\"ahnt wird. 
   
   Keplers Berechnungen\footnote{\citet[Caput XII, S. 135/6]{kepler1614DeVeroAnno}, beim julianischen Jahr \emph{29} handelt es sich wohl um einen Druckfehler, muss es vom  Kontext her nicht das Jahr \emph{39} sein?} 
   liegen -- mit Ausnahme der \emph{stella nova} -- nahe an den babylonischen Aufzeichnungen. Die erste Gro{\ss}e Konjunktion datierte er auf \emph{circa 22. Juni} und zeitgleich sei ein neuer Stern erschienen, wie damals 1604. Als weitere bedeutende Daten nennt er August und Dezember. Dazwischen in den Herbst verlegt er die Empf\"angnis Johannes des T\"aufers. Als die beiden Planeten zu Beginn des Jahres 6 v. Chr. dann die Fische verlie{\ss}en und in den Widder gingen (22. Januar), sei der Mars dazugekommen (25. Februar). Dann wanderten alle (M\"arz) an den Taghimmel, d.h. die Sonne ging durch sie hindurch und ebenfalls seien Venus und Merkur dazugetreten. Zu dieser Konjunktion aller Wandelsterne verortet er Mari\"a Verk\"undigung. Zu Beginn des folgenden Jahres 
   seien die Magier dann nach Jerusalem gekommen.

   \section{Keplers Stern in moderner Rezeption}
   
   Vermutlich, weil er fehlerhafterweise Keplers Nova \"uberlas\footnote{\citet{burke1937kepler}.}, legte der Orientalist und protestantische Bischof von Seeland, M\"unter 1821 die Hypothese vor, der Stern von Bethlehem sei einfach die dreifache Gro{\ss}e Konjunktion von 7 v. Chr. gewesen. Der englische Astronom Pritchard stellte dann 1856 neue Berechnungen vor und verlegte die Zeitpunkte der Konjunktionen auf 29.5., 29.9. und 4.12. im Jahr 7 v. Chr. und bemerkte, zum letzten Datum h\"atten beide Planeten nicht als einziger Stern wahrgenommen werden k\"onnen, denn ihr Abstand habe ca. $1^\circ$ betragen (das ist etwa der doppelte scheinbare Monddurchmesser)\footnote{ebd.}. 
   
      \subsection{Keplers Hypothese bei Ferrari d'Occhieppo}
   Der \"osterreichische Astronom \cite{ferrari1969stern}\footnote{sein Buch gibt es in Auflagen bis 2003, eine pr\"agnante Zusammenfassung gibt \citet{ferrari1978star}.} \"ubernahm M\"unters Lesung Keplers Hypothese und bezog dann Mt 2.2, \emph{wir haben seinen Stern aufgehen sehen}, auf den von den Magiern in Babylon beobachteten Aufgang des Planetenpaares im Osten, als die Sonne im Westen in der Mitte des Septembers 7 v. Chr. unterging. Der Saturn bei weitem \"uberstrahlende Jupiter galt bei den Babyloniern als K\"onigsstern und Stern des obersten Gottes Marduk. 
   Weiterhin, so Ferrari d'Occhieppo, 
   habe Saturn als Stern der Juden gegolten und das Sternbild der Fische verweise auf die westliche Gegend, die Magier h\"atten also gewusst, dass sie ins j\"udische K\"onigreich des Herodes reisen mussten, um den am Sternhimmel deutlich angek\"undigten neuen K\"onig zu sehen.    
   Den Aufbruch der Magier von Jerusalem (Mt 2.10: \emph{als sie den Stern sahen, wurden sie von sehr gro{\ss}er Freude erf\"ullt}) verlegt er auf die Abendd\"ammerung des 12. oder 13. Nov., als die beiden Planeten im S\"uden sichtbar und beide station\"ar wurden. Nach ihren Vorausberechnungen habe der Abstand nur 3 Bogenminuten betragen (etwa ein Zehntel des scheinbaren Monddurchmessers). Insbesondere ergibt sich so ein \emph{terminus ante quo} f\"ur die Geburt Jesu. Nachdem die Magier in Bethlehem angekommen waren und das Planetenpaar bereits untergegangen war, habe der \emph{Zodiakallichtkegel}\footnote{im Sonnensystem liegen nicht nur die Planeten zusammen mit der Sonne ann\"ahernd in einer Ebene, dort befinden sich auch zahlreiche Asteroiden und kleinere Himmelsk\"orper, bis hin zu Staubk\"ornern. Diese werden auch von der Sonne bestrahlt und k\"onnen als n\"achtliches Lichtband entlang der Ekliptik wahrgenommen werden, das zur Sonne hin kegelartig heller wird.} weiterhin \"uber Bethlehem gestanden und darauf bezieht er Mt. 2.9, \emph{dort blieb er [der Stern] stehen}. 
   
   \subsection{Kritischer Diskurs um Occhieppos Hypothese}

   Nach den beiden evangelischen Neutestamentlern \citet[S. 150]{theissen2013historische} \emph{halten es viele Exegeten f\"ur absurd, nach einem historischen Kern hinter Mt 2 zu suchen}, u.a. weil es \emph{viele religionsgeschichtliche Parallelen} zu dieser Erz\"ahlung gebe. 
%
   Auch der katholische Neutestamentler \citet{gaechter1970rezension} kritisiert:  \emph{Die exegetischen Bemerkungen sind bisweilen anfechtbar [...] unannehmbar ist es, wenn der Verf. im Stillstehen ein astronomisches Ph\"anomen sehen will}. Im Gegensatz dazu haben die evangelischen Neutestamentler \citet{strobel1987weltenjahr} und  \citet{stuhlmacher2006geburt} Occhieppos Hypothese im Wesentlichen \"ubernommen. 
   
   Obschon in der babylonischen Mythologie die Zuordnung des Sternbilds der Fische zum j\"udischen Land oder dem westlichen Land und von Saturn als Sinnbild der Juden \citep[S. 105 -- 109]{rosenberg1972star} problematisch erscheint (Saturn galt bei den Babyloniern u.a. wegen seiner langsamen Bewegung als die \emph{Sonne der Nacht}  und stand auch f\"ur Gerechtigkeit), so weist Rosenberg\footnote{\citet[S. 105 -- 109]{rosenberg1972star}.} dennoch auf alte j\"udische astrologische \"Uberlieferungen hin, nach denen Gro{\ss}e Konjunktionen, insbesondere in den Fischen auf gro{\ss}e Ereignisse und insbesondere auf den Messias hindeuten. Vermutlich waren diese \"Uberlieferungen Kepler bekannt.
   Aufgrund dieser eben genannten Problematik weist dann auch der Religionswissenschaftler und Naturwissenschaftshistoriker \cite{papke1999das} Occhieppos Hypothesen zur\"uck, zwar werde Saturn in Amos 5,26, \emph{ihr werdet den Sakkut als euren K\"onig vor euch hertragen m\"ussen und den Kewan, euren Sterngott, eure G\"otter, die ihr euch selber gemacht habt}, mit seinen aram\"aischen Bezeichnungen\footnote{tats\"achlich seien \begin{cjhebrew}\cjRL{skwt}\end{cjhebrew} und \begin{cjhebrew}\cjRL{kywn}\end{cjhebrew} \emph{hapax legomena}, so \citet{gevirtz1968new}, er gibt dennoch eine l\"angere Liste von Literatur, die diese auf Saturn beziehen.}  erw\"ahnt, aber offensichtlich stehe er eben nicht f\"ur Israel, sondern f\"ur seinen Abfall. Auch entspr\"ache, so \citet[S. 29]{papke1999das}, es \emph{eindeutig nicht der Realit\"at}, dass der viermal  in Mt 2 verwandte Singular, \emph{Stern} die beiden Planeten bezeichne. Dieser letzten Kritik schlie{\ss}t sich u.a. der Philosoph und Altphilologe, \citet{koch2006stern} an. 

   Im Gegensatz dazu \"ubernimmt der Astronom \citet[S. 254 -- 313]{schmidt2005stern} vollumf\"anglich Occhieppos
   Hypothesen, wohingegen der Mathematikhistoriker Neugebauer\footnote{er verweigerte den erneuerten Beamteneid auf Hitler und emigrierte 1934.} etwas kryptisch Bedenken anmeldet, es sei \emph{offenbarer Unsinn}, wenn Forscher, \emph{die nicht an Wunder glauben, die durch Naturereignisse nach Herzenslust ersetzen}\footnote{\citet[S. 608]{neugebauer1975history}: \emph{... but obvious nonsense. Moderm scholars [...]
do not believe in the miracles but replace them to their hearts' content by possible
natural events.}.}.
   
   Eine offensichtliche, weitere Kritik ergibt sich aus dem im Abschnitt \ref{scn:omina} vorgestellten babylonischen Archiv: Die dreifache Gro{\ss}e Konjunktion erhielt offenbar keinerlei Aufmerksamkeit. Warum sollten sich also die Magier daf\"ur interessiert haben? Tats\"achlich l\"asst sich dieses Argument aber auch umkehren: H\"atte sich eine breitere babylonische Wissenschaftskultur f\"ur diese Konjunktion interessiert, w\"are dann Jerusalem nicht von Magiern \"uberflutet worden?

   Eine umfangreiche, ausf\"uhrliche und kritische \"Ubersicht \"uber \"altere und zeitgen\"ossische astronomische Hypothesen gibt \citet[S. 116 -- 228]{koch2006stern} zusammen mit der von ihm favorisierten Hypothese, der Stern von Bethlehem sei allein Venus gewesen\footnote{als innerer Planet ist Venus bei ihrem heliakalischen Aufgang (d.h. Erstaufgang morgens im Osten, kurz vor der Sonne, sie steht dann, anders als die \"au{\ss}eren Planeten beim heliakalischen Aufgang zwischen Erde und Sonne -- nur die inneren Planeten kennen auch einen Abendaufgang) der einzige Planet (au{\ss}er Merkur), der dann bereits r\"uckl\"aufig ist, d.h. sie zieht nicht (zwar etwas langsamer) mit der Sonne entgegen der Hintergrundbewegung des Sternhimmels, sondern mit der Hintergrundbewegung und etwas schneller, so deutet \citet[S. 98 -- 101]{koch2006stern} Mt 2,9 \emph{der Stern ... zog vor ihnen her.} Allerdings wird sie erst nach etwa 2 Wochen station\"ar (danach entfernt sie sich aber weiterhin von der Sonne, zieht aber dann mit der Sonne entgegen dem Sternenhintergrund). 
   %
   %
   Warum \citet[S. 102 f.]{koch2006stern} \emph{zwischen dem heliakischen Aufgang dieses Sterns und seinem Stillstand nur wenige Tage} annimmt, kann ich, nachdem ich die zeitlichen Abst\"ande in einen Zeitraum von mehr als 8 Jahren (das entspricht in etwa dem \glqq Saroszyklus\grqq~der Venus, \citet[S. 156]{neugebauer1975history}) \"uberpr\"ufte, nicht nachvollziehen.}.
   
   Die ungebrochene Aktualit\"at der Thematik spiegelt u.a. eine Konferenz \glqq The Star of Bethlehem: Historical and Astronomical Perspectives\grqq~(2014)\footnote{\cite{van2015star}.} an der Universit\"at Groningen wider, in der Astronomen, Exegeten und Historiker in 
   Diskurs traten. In einem Zentrum stand Keplers Hypothese, auch in Ferrari d'Occhieppos Lesung. Ein zweites Zentrum -- der eigentliche Anlass der Konferenz -- bildete die Erkl\"arung des Astronomen Michael Molnar\footnote{\cite{molnar1999star}, auf dieser Konferenz wies \cite{de20156} auf einen breiteren, aber bislang weniger wahrgenommenen geschichtlichen Kontext zu Molnars Erkl\"arung hin.} als eine doppelte, heliakalische Bedeckung des Jupiters durch den Mond am 20. M\"arz und 17. April 6 v. Chr. im Widder, die im Einklang mit zeitgen\"ossischen astrologischen Vorstellungen auf eine K\"onigsgeburt bei den Juden hingewiesen habe. 

%
%

   \section{Epilog: Grenzen der Naturwissenschaft  und Anfragen an die Theologie}
   
   Neben der Verortung des Todes Jesu unter Pontius Pilatus kennt das christliche Glaubensbekenntnis keine weiteren zeitlichen Verortungen. Selbst die Tradition kann keine eindeutige Antwort zum Geburts- und Todesdatum Jesu geben. Was w\"are, 
   wenn Keplers Berechnung im Wesentlichen die Erz\"ahlung um den Stern von Bethlehem historisch verortbar machte und dieser damit insbesondere kein \glqq Wunder\grqq~mehr w\"are? Tats\"achlich scheinen moderne Naturwissenschaftler weniger Schwierigkeiten mit Wundern zu haben, als Theologen, f\"ur die Wunder Durchbrechen von Naturgesetzen sind und oftmals aus diesem Grunde nicht sein k\"onnen. F\"ur Naturwissenschaftler gibt es nur Modelle und sie wissen: \emph{alle Modelle sind falsch, einige aber n\"utzlich}\footnote{\cite{box1976}}. 
   
 Im Rahmen dieser Fragestellung ergab sich als Korollar aus der Betrachtung der naturwissenschaftlichen \"Uberlegungen ein ggf. unerwarteter Preis, den die Naturwissenschaft 
   zahlen muss:
   Ihr erscheinen 
   dadurch Grenzen gesetzt, dass der Kosmos so eingerichtet ist, dass wohl keine  Harmonien gefunden werden k\"onnen, komplexere empirische Gr\"o{\ss}en w\"aren damit immer inkommensurabel, so wie die Seitenl\"angen des Pentagramms: Die o.g. numerischen Methoden erfassen also letztlich nicht die Wirklichkeit.  
   Welches Gottesbild steht hinter einem solch unharmonischen Kosmos?
   
   In diesem Beitrag wurde auch deutlich, dass die Astrologie zusammen mit der Astronomie zun\"achst ein Zweig der Naturwissenschaft war. F\"ur die Astronomie konnte erfolgreich der Nachweis gef\"uhrt werden, dass Himmelsereignisse durchaus gewisse Periodizit\"aten aufweisen, eben nur \emph{gewisse}, eingeschr\"ankt u.a. durch die letztliche Inkommensurabilit\"at. Der Nachweis einer Korrelation mit Ereignissen auf der Erde scheiterte offenbar, so dass die Astrologie den Status einer Wissenschaft verlor und zum Aberglauben wurde. Ist die Erz\"ahlung um den Stern von Bethlehem historisch verortbar, so war in diesem konkreten Ausnahmefall die Astrologie offenbar kein Aberglaube, sondern korrekt, ja im Vollsinn des Wortes \emph{zielf\"uhrend}. Aber auch anderenfalls, wenn die Erz\"ahlung rein kerygmatisch w\"are, so geh\"orte die zielf\"uhrend angewendete Astrologie zum Kerygma. Wiederum: Welches Gottesbild steht dahinter? Kann Astrologie also, damals wie heute \emph{Zeichen der Zeit}\footnote{Gaudium et Spes 4.} sein?


      \end{document}